\documentclass[aps,prb,twocolumn,reprint,superscriptaddress,floatfix]{revtex4-2}

\usepackage[T1]{fontenc}
\usepackage{lmodern}
\usepackage{graphicx}
\usepackage{amsmath}
\usepackage{amssymb}
\usepackage{braket}
\usepackage{hyperxmp}

\usepackage{tikz}
\usetikzlibrary{svg.path}
\definecolor{orcidlogocol}{HTML}{A6CE39}
\tikzset{
  orcidlogo/.pic={
    \fill[orcidlogocol] svg{M256,128c0,70.7-57.3,128-128,128C57.3,256,0,198.7,0,128C0,57.3,57.3,0,128,0C198.7,0,256,57.3,256,128z};
    \fill[white] svg{M86.3,186.2H70.9V79.1h15.4v48.4V186.2z}
                 svg{M108.9,79.1h41.6c39.6,0,57,28.3,57,53.6c0,27.5-21.5,53.6-56.8,53.6h-41.8V79.1z M124.3,172.4h24.5c34.9,0,42.9-26.5,42.9-39.7c0-21.5-13.7-39.7-43.7-39.7h-23.7V172.4z}
                 svg{M88.7,56.8c0,5.5-4.5,10.1-10.1,10.1c-5.6,0-10.1-4.6-10.1-10.1c0-5.6,4.5-10.1,10.1-10.1C84.2,46.7,88.7,51.3,88.7,56.8z};
  }
}

\newcommand\orcidicon[1]{\href{https://orcid.org/#1}{
\begin{tikzpicture}[yscale = -1, scale =0.030,  transform shape, baseline={(0,-8pt)}]
\pic{orcidlogo};
\end{tikzpicture}
}$\!\!$}

\usepackage[breaklinks]{hyperref}
\hypersetup{
	colorlinks=true,
	linkcolor=blue,
	citecolor=blue,
	urlcolor=blue,
	pdftitle= {High magnetic field spin-valley-split Shubnikov--de Haas oscillations in a WSe$_2$ monolayer},
	pdfauthor = {Banan Kerdi, Mathieu Pierre, Robin Cours, B\'{e}n\'{e}dicte Warot-Fonrose, Michel Goiran, Walter Escoffier},
	pdfkeywords = {Transition metal dichalcogenides (TMDCs), WSe$_2$ monolayers, magneto-transport, high magnetic field physics, Zeeman effect, Land\'{e} factor, Landau levels, Shubnikov--de Haas oscillations},
	pdfpagemode=UseNone,
}

\begin{document}

\title{High magnetic field spin-valley-split Shubnikov--de Haas oscillations in a WSe$_2$ monolayer}

\author{Banan Kerdi\,\orcidicon{0000-0002-6015-1850}}
\email{banan.kerdi@lncmi.cnrs.fr}
\affiliation{LNCMI, Universit\'e de Toulouse, CNRS, INSA, UPS, EMFL, 31400 Toulouse, France}

\author{Mathieu Pierre\,\orcidicon{0000-0001-5274-4309}}
\affiliation{LNCMI, Universit\'e de Toulouse, CNRS, INSA, UPS, EMFL, 31400 Toulouse, France}

\author{Robin Cours}
\affiliation{CEMES, Universit\'e de Toulouse, CNRS, 31055 Toulouse, France}

\author{B\'{e}n\'{e}dicte Warot-Fonrose\,\orcidicon{0000-0001-9829-9431}}
\affiliation{CEMES, Universit\'e de Toulouse, CNRS, 31055 Toulouse, France}

\author{Michel Goiran}
\affiliation{LNCMI, Universit\'e de Toulouse, CNRS, INSA, UPS, EMFL, 31400 Toulouse, France}

\author{Walter Escoffier}
\email{walter.escoffier@lncmi.cnrs.fr}
\affiliation{LNCMI, Universit\'e de Toulouse, CNRS, INSA, UPS, EMFL, 31400 Toulouse, France}

\begin{abstract}
We study Shubnikov--de Haas oscillations in a p-type WSe$_2$ monolayer under very high magnetic field. The oscillation pattern is complex due to a large spin and valley splitting, in the non-fully-resolved Landau level regime. 
Our experimental data can be reproduced with a model in which the main parameter is the ratio between the Zeeman energy  and the cyclotron energy. 
The model takes into account the Landau levels from both valleys with the same Gaussian broadening, which allows to predict the relative amplitude of the resistance oscillation originating from each valley.
The Zeeman energy is found to be several times larger than the cyclotron energy. 
It translates into a large and increasing effective Land\'{e} factor as the hole density decreases, in the continuity of the values reported in the literature at lower carrier density.
\end{abstract}

\maketitle

\section{Introduction}

Single atomic layers of semiconducting transition metal dichalcogenides (TMDCs) have received much attention over the last decade due to their promising characteristics for two-dimensional-based (2D) optoelectronic devices \cite{Splendiani2010, Mak2010,Yin2012,Lopez-Sanchez2013} and their potential for the development of valleytronics physics \cite{Xiao2012}. The crystal structure of these 2D materials, composed of one transition metal atom (W, Mo) and two chalcogen atoms (S, Se) per unit cell arranged in a honeycomb lattice, leads to a direct band gap in the visible range, located at the corner (K and K$'$ points) of the first Brillouin zone \cite{Splendiani2010, Mak2010,Zhao2013}. The lack of inversion symmetry of the crystal structure, combined with strong spin-orbit coupling due to the heavy transition metal atoms, translate into coupled valley and spin degrees of freedom of the charge carriers, later referred to as the $\ket{\mathrm{K}\uparrow}$ and $\ket{\mathrm{K'}\downarrow}$ states. 
The presence of an out-of-plane magnetic field not only turns the energy spectrum into discrete Landau levels (LL), separated by the cyclotron energy $E_c$, but also lifts the spin/valley degeneracy of the $\ket{\mathrm{K}\uparrow}$ and $\ket{\mathrm{K'}\downarrow}$ states through Zeeman energy $E_z=g^* \mu_\mathrm{B} B$ where $\mu_\mathrm{B}$ is the Bohr magneton and $g^*$ is the effective Land\'{e} factor. In TMDC monolayers, experimental studies point to a very large value of $g^*$ since the spin, orbital, and lattice Zeeman effects add up to each other \cite{Pisoni2018b, Gustafsson2018, Movva2017}. The first effect refers to the interaction between the magnetic field and the electron or hole spin, the second is linked to the spin-orbit coupling, and the last alludes to the opposite Berry curvatures in the non-equivalent K and K$'$ valleys \cite{Aivazian2015}. 
While the effective Land\'{e} factor of excitons in TMDC monolayers has been measured by several groups \cite{Mitioglu2015, Stier2016, Koperski2018, Goryca2019}, it would be desirable to obtain this fundamental parameter for the conduction and valence bands independently. 
$g^*$ is expected to depend on the chemical composition and the number of layers of the TMDC, on the nature (electron or hole) of the charge carriers, as well as the carrier density. 
In this work, we focus on its determination for holes in WSe$_2$ monolayers, which we extend to high carrier concentration thanks to  a very high magnetic field.

\begin{figure}[tp]
\centering
\includegraphics[width=1\linewidth]{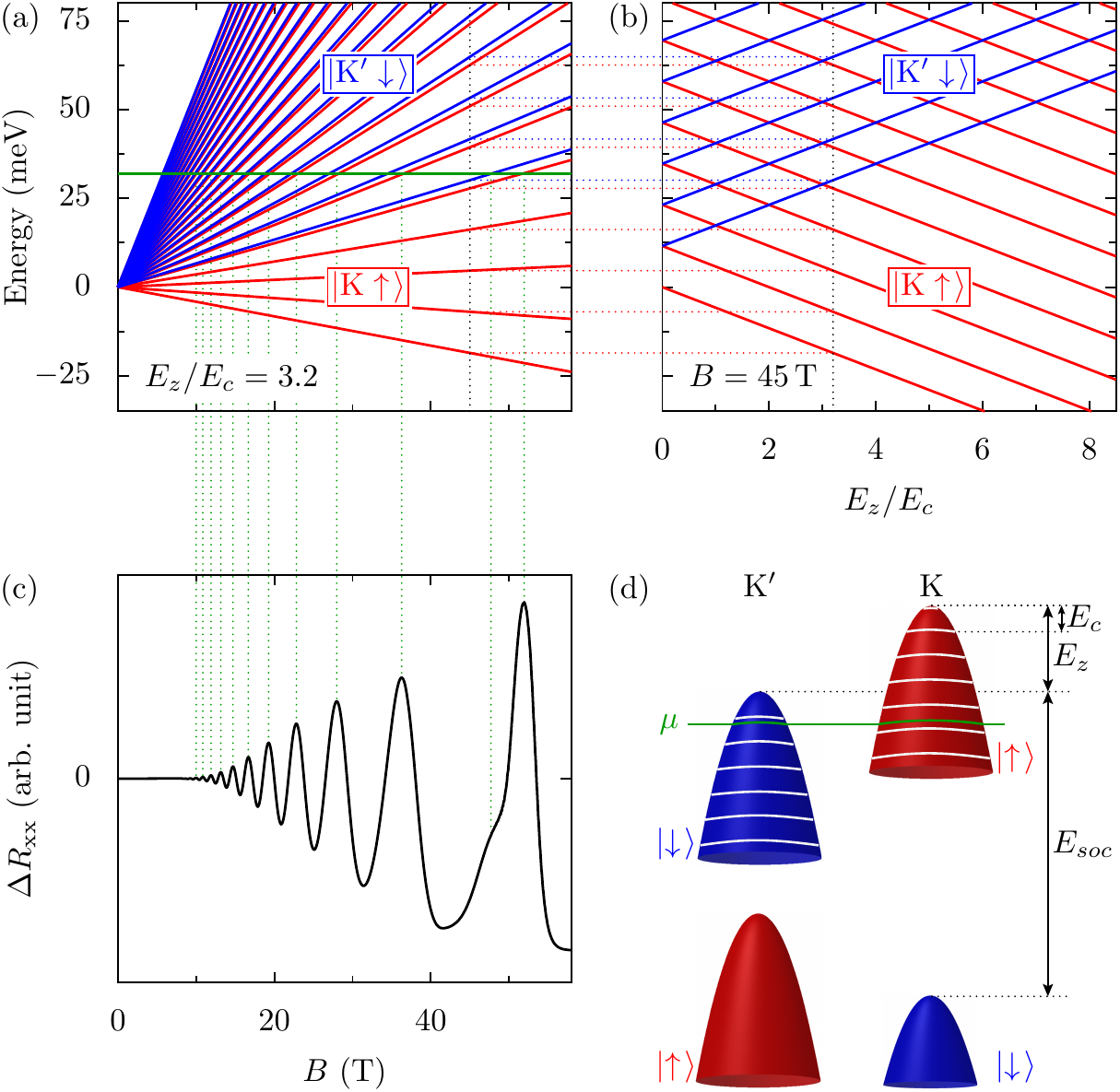}
\caption{\label{Fig.1} Energy of LLs as a function of (a) magnetic field and (b) $E_z/E_c$ where the red and blue lines represent the $\ket{\mathrm{K}\uparrow}$ and $\ket{\mathrm{K'}\downarrow}$ states, respectively. (c) Simulated magneto-resistance where both a large LL broadening and a non-integer ratio $E_z/E_c=3.2$ have been chosen to produce an example of complex oscillations. The simulation is performed at constant chemical potential corresponding to the solid green line depicted in panel (a).
A maximum of resistance is expected each time the chemical potential crosses a LL. (d) Simplified valence band of WSe$_2$ monolayers under magnetic field. The valleys at points K and K$'$ of the first Brillouin zone are made of two spin-split parabolic bands. For a given spin sign, the energy difference between the two bands arises from the spin-orbit coupling (SOC), with $E_\mathrm{soc} = 466$\,meV \cite{Liu2013, Le2015}.  The white lines indicate the energies of the Landau levels for $E_z/E_c = 3.2$. They are shown only for the two upper bands, where the chemical potential (solid green line) is located in the explored range of hole densities.}
\end{figure}

In general, the ratio $E_z/E_c$, from which $g^*$ is extracted, can take any value, leading to  non-equidistant Landau levels and  complex Shubnikov--de Haas  (SdH) oscillations in a magneto-transport experiment.
Figure~\ref{Fig.1} shows the Landau level spectrum as a function of magnetic field in the case of a large spin-valley splitting (Fig.~\ref{Fig.1}a) and its evolution with $E_z/E_c$ (Fig.~\ref{Fig.1}b). When this ratio is not an integer, the non-equidistant Landau levels translate into non $1/B$-periodic quantum oscillations, as illustrated in Fig.~\ref{Fig.1}c. 
In 2D electron gases (2DEG) realized in standard semiconducting heterostructures, the effective Land\'{e} factor is experimentally addressed by tilting the magnetic field with respect to the sample's plane. 
Indeed, the Zeeman energy $E_z$ depends on the magnetic field intensity regardless of its orientation whereas the cyclotron energy $E_c$ is proportional to the out-of-plane projection of the magnetic field.
A coincidence angle is reached when the Zeeman energy equals the cyclotron energy, and the magneto-resistance recovers the distinctive $1/B$-periodicity of single-carrier Shubnikov--de Haas oscillations. In the case of TMDCs, however, this coincidence method fails because the spin of the charge carriers is locked to the sample's out-of-plane direction \cite{Movva2017,Zhu2011}.
Nonetheless, and beyond the  difficulty in obtaining high-mobility samples with low-resistance ohmic contacts, a few studies have reported the ratio $E_z/E_c$ based on the analysis of SdH oscillations.
In Ref.~\onlinecite{Movva2017}, an estimation of $E_z/E_c$ rounded to integer values has been extracted from  sequences of  SdH oscillation minima at odd or even filling factors. 
In Ref.~\onlinecite{Lin2019}, the identification of the transition between mixed and polarized Landau level regimes has been exploited  to deduce   an electron-density dependent $g^*$ factor in high-mobility MoS$_2$ bilayers.
Otherwise, as an alternative to magneto-transport, contactless spectroscopy using a single-electron transistor has been used to sense directly the chemical potential and infer the Landau level spectrum \cite{Gustafsson2018}.

In this work, we performed magneto-transport measurement in a WSe$_2$ monolayer under very high magnetic field, allowing for partial Landau level resolution in samples with reasonable mobilities fabricated without complicated contact engineering. We analyze the complex SdH oscillation pattern using a model where $E_z/E_c$ is the main fitting parameter. It takes into account all the $\ket{\mathrm{K}\uparrow}$ and $\ket{\mathrm{K'}\downarrow}$ Landau levels at once with the same Gaussian broadening. This novel approach is particularly helpful to analyze oscillations with non-fully-resolved spin and valley splitting.

\begin{figure}[htp]
\centering
\includegraphics[width=1\linewidth]{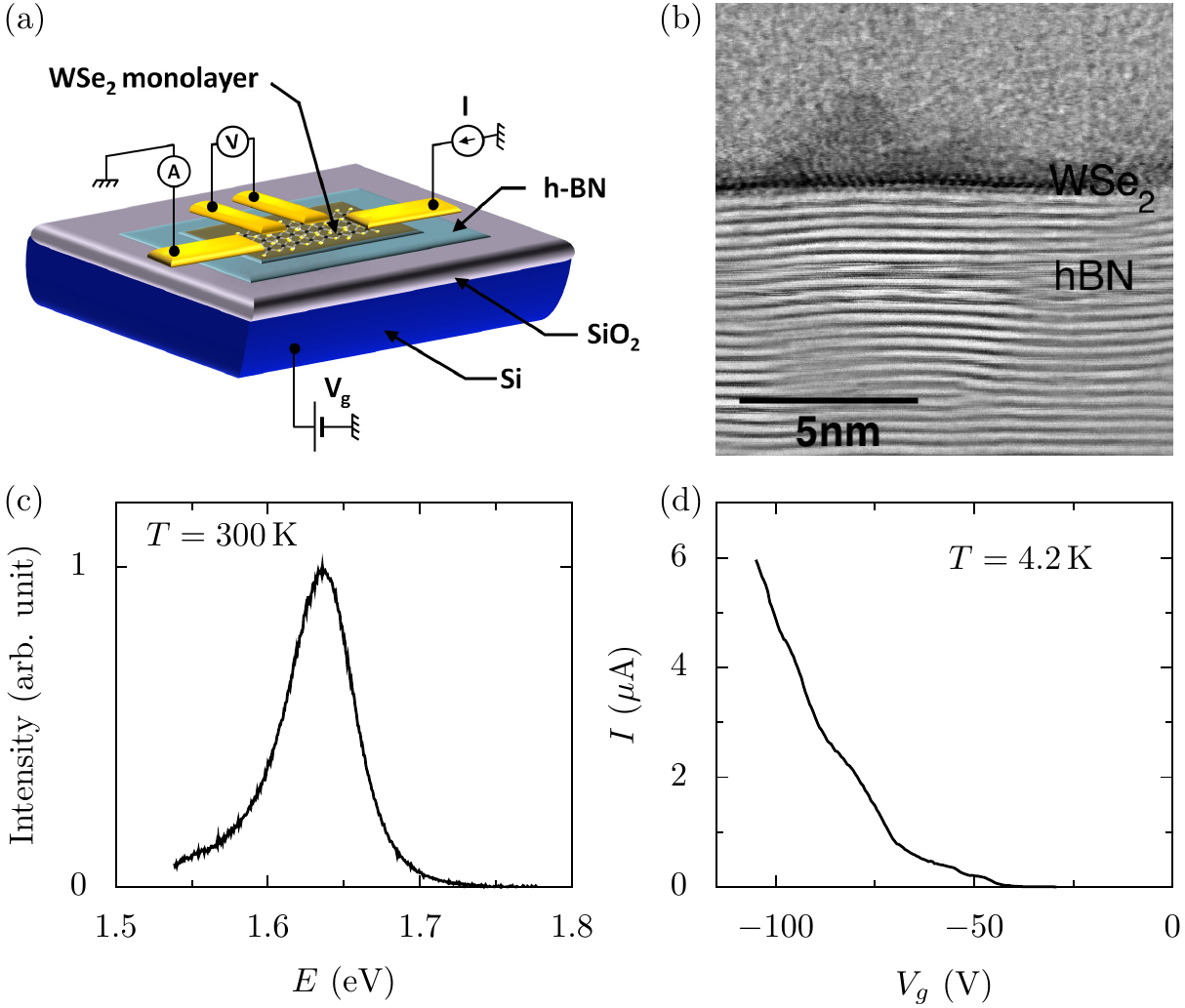}
\caption{\label{Fig.2} (a) Sketch of the BN/WSe$_2$ heterostructure. (b) Bright-field scanning transmission electron microscopy (BF-STEM) image realized after the measurement campaign confirming a WSe$_2$ monolayer as the conducting channel. The h-BN substrate appears as dark/bright fringes whereas the WSe$_2$ layer consists of dark spots surrounded by brighter ones. The interatomic distances between W and Se atoms correspond to the expected ones.
(c) The photoluminescence spectrum with $\lambda_\mathrm{exc} = 633$\,nm is characteristic of a WSe$_2$ monolayer. (d) Current flowing in the device for $V_{ds}=100$\,mV versus the back-gate voltage.}
\end{figure}

\section{Experiments}

A single atomic layer of WSe$_2$ is exfoliated from the bulk material \cite{ref_fabricant} using the micro-mechanical cleavage method and transferred onto a stamp of PDMS \cite{Castellanos_Gomez_2014}. This flake is deposited under optical microscope control onto a 20-nm-thick flake of Boron Nitride (BN) pre-deposited on a standard Si/SiO$_2$ substrate with $d_{\mathrm{SiO}_2} = 280$\,nm. Electron beam lithography with a 495\,K (100\,nm)/950\,K (80\,nm) PMMA bilayer mask  followed by platinum sputtering (10\,nm) and gold thermal evaporation (50\,nm), is realized to fabricate the electrodes, as sketched in Fig.~\ref{Fig.2}a. The WSe$_2$ flake is later confirmed to be a monolayer by photoluminescence (Fig.~\ref{Fig.2}c) and bright-field scanning transmission electron microscopy (Fig.~\ref{Fig.2}b). 
In this transistor configuration, the back-gate voltage ($V_g$) applied between the sample and the substrate sets the hole density.

\begin{figure}[htp]
\centering
\includegraphics[width=1\linewidth]{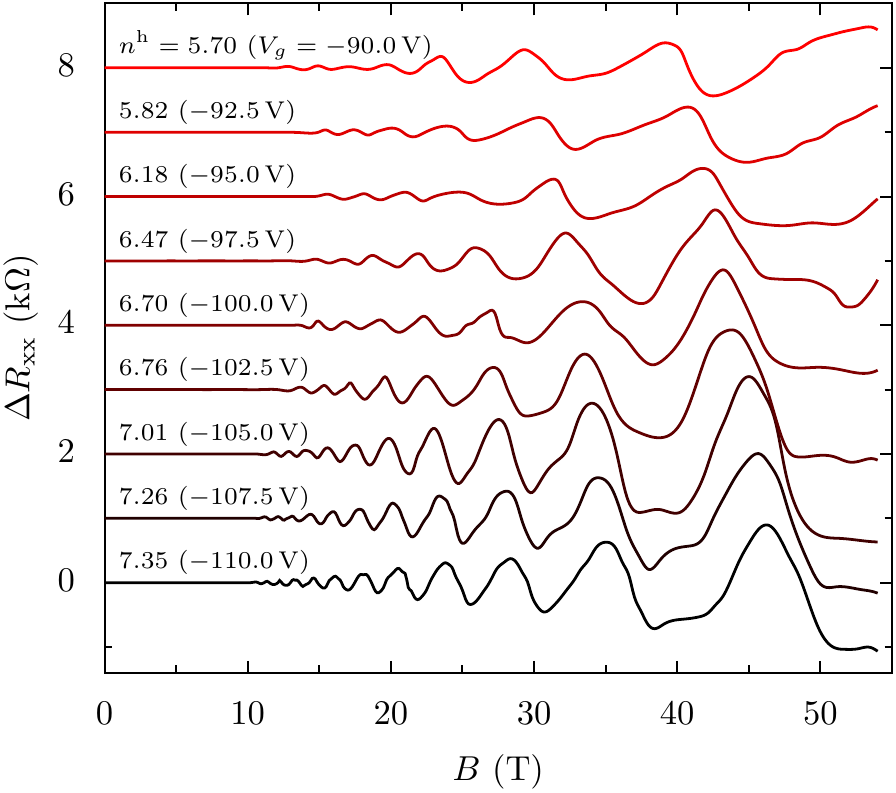}
\caption{\label{Fig.3} High field magneto-resistance of the sample at $T=4.2$\,K, after background removal, for various values of the carrier density $n^\mathrm{h}$ (given in unit of $10^{12}\,\mathrm{cm}^{-2}$) ranging from 5.70 to $7.35 \times 10^{12}\,\mathrm{cm}^{-2}$. The corresponding back-gate voltage $V_g$ is given in parentheses. An offset of $1\,$k$\Omega$ is set between consecutive curves for clarity. The strong deviations from a $1/B$-periodic oscillatory behavior are indicative of an additional Zeeman energy scale competing with the cyclotron energy. See also Supplemental Material I where the same dataset is plotted versus filling factor.}
\end{figure}

We performed transport measurements under pulsed magnetic field up to 54\,T at low temperature (4.2\,K) for different values of the back-gate voltage. 
A constant DC current of $1\,\mu$A is passed through the sample while the longitudinal voltage is measured during a pulse of magnetic field with total duration  $\sim500$\,ms.
The magneto-resistance displays large-amplitude oscillating features on top of a smooth background, subtracted to the data to obtain the oscillatory part $\Delta R_{\mathrm{xx}}(B)$ of the signal only, as shown in Fig.~\ref{Fig.3}. 
For all gate voltages, the oscillations exhibit a complex pattern, requiring an analysis beyond spin-degenerate single-band $1/B$-periodic SdH oscillations. 
The large spin/valley splitting is responsible for this effect, as explained in the Introduction.

\section{Modeling and simulations}

To extract the ratio $E_z/E_c$, we performed numerical simulations of the oscillatory part of the resistance $\Delta R_{\mathrm{xx}}(B)$. We start with the LL structure of WSe$_2$, adapted from the massive Dirac fermion dispersion relation \cite{Rose2013, Li2013, Kormanyos2015}, which provides the set of energies $E_{N,s}$ at a given magnetic field $B$.
\begin{equation}
E_{N,s}=-N \hbar \omega_c - s g^* \mu_\mathrm{B} B,
\label{eq.1}
\end{equation}
where $\omega_c=eB/m^*$ is the cyclotron energy, $g^*$ is the effective Land$\'{e}$ factor, $N=0, 1, 2, ...$ is the LL index, and $s=\pm \frac{1}{2}$ stands for the coupled spin/valley degree of freedom, respectively $\ket{\mathrm{K}\uparrow}$ and $\ket{\mathrm{K'}\downarrow}$. We set $s=+\frac{1}{2}$ for $N=0$ to satisfy energy minimization argument \cite{Wang2017}. Under quantizing magnetic field and assuming a finite energy relaxation time, the density of states can be described by a sum of area-normalized Gaussian functions centered at the energy of the LLs $E_{N,s}$,
\begin{equation}
\mathrm{DoS}(E,B)=\sum\limits_{N,s} \frac{eB}{h} \times \frac{1}{\sqrt{2\pi} \Gamma (B)} \times \exp\left(\frac{-\left(E-E_{N,s}\right)^2}{2\Gamma^2(B)}\right),
\label{eq.2}
\end{equation}  
where $\Gamma (B) =\frac{\hbar e}{m^*} \sqrt{\frac{2 B}{\pi \mu^\mathrm{h}}}$  with $\mu^\mathrm{h}$ the hole mobility at zero magnetic field \cite{Ando1974}. The pre-factor $eB/h$ accounts for the orbital degeneracy of the LLs. We assume that the charge density $n^\mathrm{h}$ remains constant in the device, set by the fixed back-gate voltage $V_g$, while the chemical potential changes according to the variation of the density of states at the Fermi energy induced by the magnetic field. For each value of the magnetic field, the chemical potential $\mu (B)$ is obtained numerically by solving
\begin{equation}
n^\mathrm{h}=\int\limits_{-\infty}^{+\infty} \mathrm{DoS}(E,B) \times f \left(E, \mu(B), T\right) \times dE,
\label{eq.3}
\end{equation}
i.e., when  the cumulative orbital degeneracy of the occupied LLs reaches the hole carrier density. Here, $f(E, \mu, T)= \frac{1}{1+\exp{\frac{\left(E-\mu\right)}{k_\mathrm{B}T}}}$ is the Fermi-Dirac distribution function. The conductivity $\sigma_{\mathrm{xx}}(B)$ is calculated from \cite{Gerhardts2008}
\begin{multline}
\sigma_{\mathrm{xx}}(B) = \frac{e^2}{h} \sum\limits_{N,s} \int\limits_{-\infty}^{+\infty} \left(N+\frac{1}{2}\right) \\
\times \left[\exp\left(\frac{-\left(E-E_{N,s}\right)^2}{2\Gamma^2(B)}\right)\right]^2 \times \left[\frac{\partial f (E, \mu, T)}{\partial E}\right] 
 \times dE. 
\label{eq.4}
\end{multline}
Here, the conductivity of the system is interpreted within the two-fluid model without interaction between the charge carriers belonging to different valley/spin indices. The contributions of each quantum state to the conductivity add therefore independently. We emphasize that this model is certainly oversimplified since it cannot reproduce LL anti-crossing as investigated in Refs.~\onlinecite{Lin2019, Pisoni2018a} for MoS$_2$ and Ref.~\onlinecite{Gustafsson2018} for WSe$_2$. The longitudinal resistivity $\rho_\mathrm{xx}$ is computed by inverting the conductivity tensor, where the product $\sigma_\mathrm{xx} \times \rho_\mathrm{xy} \ll 1$ can be expanded in a Taylor series when the chemical potential is located in between two successive LLs. We obtain
\begin{equation}
\rho_\mathrm{xx}=\sigma_\mathrm{xx}\times\rho_\mathrm{xy}^2+\sigma_\mathrm{xx}^3\times \rho_\mathrm{xy}^4 + \dots,
\label{eq.5}
\end{equation}
where the Hall resistivity $\rho_\mathrm{xy}$ is computed from the relation
\begin{equation}
\rho_\mathrm{xy}=\frac{h}{e^2} \left[\int\limits_{-\infty}^{+\infty}\sum\limits_{N,s} \frac{dE}{\sqrt{2\pi}\Gamma (B)} \times \exp\left(-\frac{\left(E-E_{N,s}\right)^2}{2\Gamma^2(B)}\right) \right]^{-1}.
\label{eq.6}
\end{equation}

\newpage

\begin{figure}[ht]
\centering
\includegraphics[width=1\linewidth]{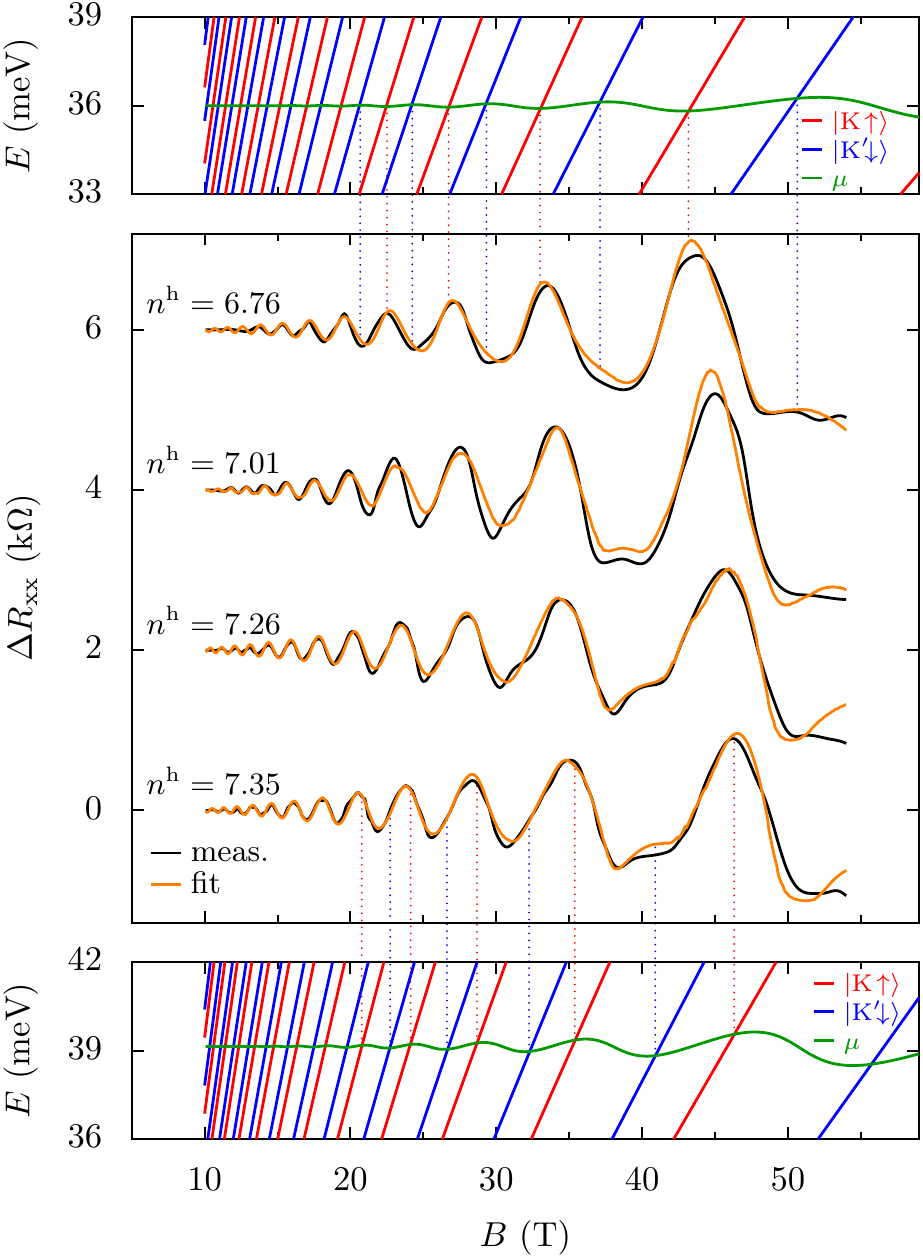}
\caption{\label{Fig.4} Middle frame: The experimental magneto-resistance (black lines) are compared to the model output (orange lines) for selected hole densities $n^\mathrm{h}$ (given in unit of $10^{12}\,\mathrm{cm}^{-2}$). Top and bottom frames: Landau level spectrum and evolution of the chemical potential $\mu (B)$ for  $n^\mathrm{h} = 6.76 \times10^{12}$\,cm$^{-2}$ and $n^\mathrm{h} = 7.35 \times10^{12}$\,cm$^{-2}$, respectively. Deviations from $1/B$-periodic oscillations appear at high field, as a direct consequence of the non-integer $E_z/E_c$ ratio. See also Supplemental Material II for additional fits of the data.}
\end{figure}

Despite some mismatch in the relative amplitude of the oscillations, probably linked to imprecise background subtraction, the overall shape of the curves is nicely reproduced without invoking any change of $E_z/E_c$ with magnetic field (see Fig.~\ref{Fig.4}).
The best fits provide the hole density $n^\mathrm{h}$, the mobility $\mu^\mathrm{h}$ and the ratio $E_z/E_c$. 
We used the same mobility for the $\ket{\mathrm{K}\uparrow}$ and $\ket{\mathrm{K'}\downarrow}$ states as we suppose that the main sources of scattering (remote charged impurities and punctual lattice defects) are spin/valley independent. 
The effective mass is set to $m^*=0.45\times m_e$ where $m_e$ is the bare electron mass (see Supplemental Material IV) and the temperature is set constant at $T=4.2$\,K. The numerical simulation also includes the optimization of the mobility edge, which sets the energy threshold between localized and extended states in LLs, as a magnetic field-dependent cut-off of the Gaussian functions in Eqs.~\eqref{eq.4} and \eqref{eq.6}. 
The extracted hole density with respect to $V_g$ is in good agreement with the plane capacitor model (see Supplemental Material III) where the gate capacitance per unit area, $C_g = \left( C_\mathrm{SiO_2}^{-1} + C_\mathrm{BN}^{-1} \right)^{-1} = 11.5\,\mathrm{nF}/\mathrm{cm}^2$, is derived from the SiO$_2$ thickness of 280\,nm and the BN thickness of 20\,nm with relative dielectric permittivity $\epsilon_{r(\mathrm{SiO_2})}=3.9$ and $\epsilon_{r(\mathrm{BN})}=3.8$, respectively.
We obtain $\mu^\mathrm{h}\sim$ 2000\,cm$^2$V$^{-1}$s$^{-1}$ for $n^\mathrm{h} \gtrsim 6.5\times10^{12}$\,cm$^{-2}$,
 with a progressive drop down to $\mu^\mathrm{h}\sim$ 1000\,cm$^2$V$^{-1}$s$^{-1}$ as the hole density decreases.  
For  $n^\mathrm{h} < 5.7 \times10^{12}$\,cm$^{-2}$, the amplitude of the oscillations is too weak with respect to the magneto-resistance background, preventing a reliable determination of $E_z/E_c$.

\begin{figure}[t]
\centering
\includegraphics[width=1\linewidth]{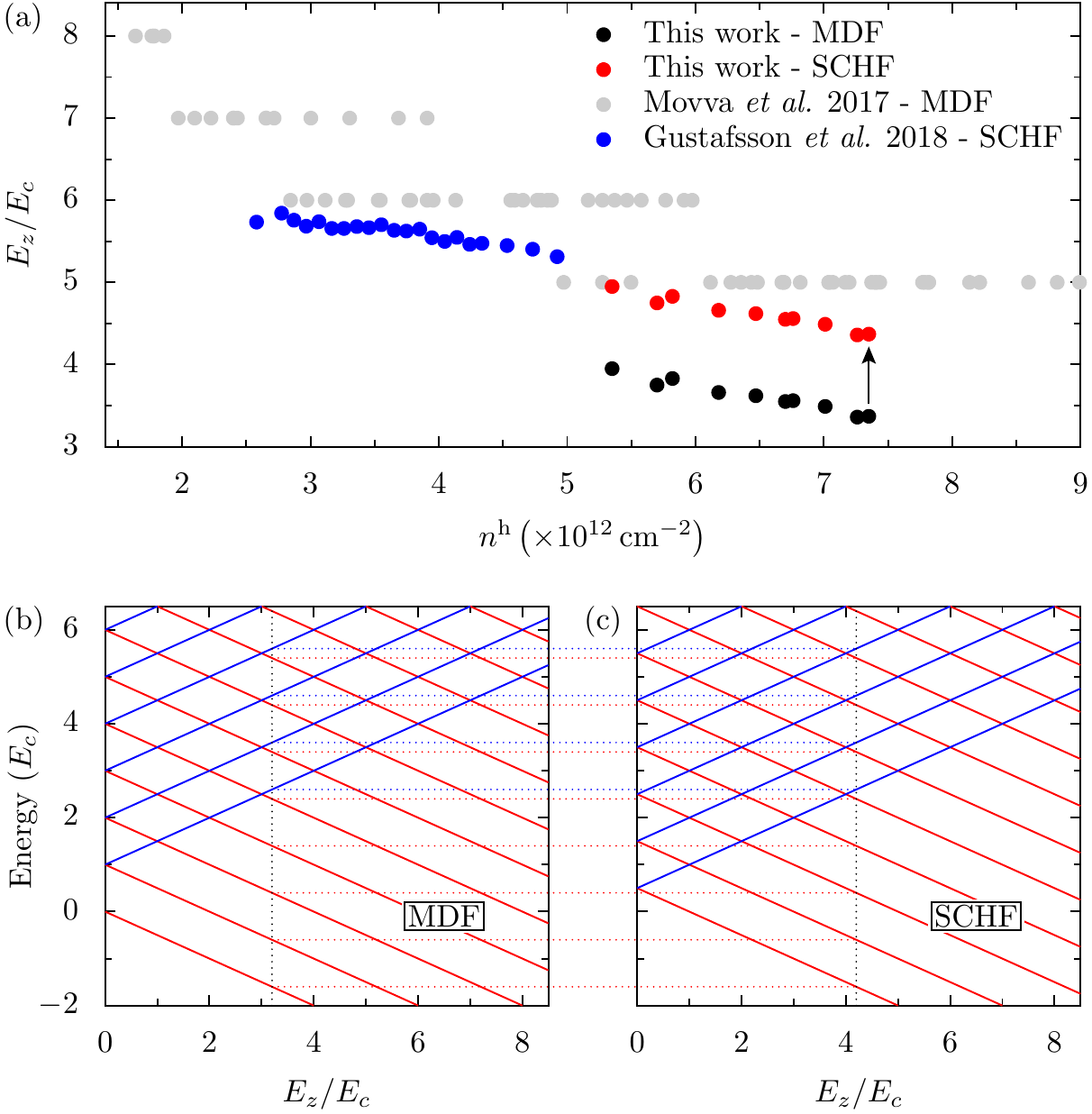}
\caption{\label{Fig.5} (a) Extracted values of $E_z/E_c$ as a function the hole density. The black dots are the values obtained assuming the Landau level structure of Massive Dirac Fermions (MDF). By up-shifting these values by 1, we obtain the values corresponding to the LL structure of Schr\"odinger Fermions (SCHF). $E_z/E_c$ increases as the carrier density reduces, suggesting the reinforcement of the Coulomb interactions. The $E_z/E_c$ ratio is compared to the values found in the literature (blue and gray dots), where either the MDF or SCHF model was considered (see main text). (b),  (c) Comparison of LL energies in cyclotron energy unit as a function of $E_z/E_c$ for the (b) MDF and (c) SCHF models. This figure demonstrates that their LL structure is identical under an up-shift of $E_z/E_c$ by~1.}
\end{figure}

\section{Discussion}

Our simulations suggest that the ratio $E_z/E_c$ is comprised between 3 and 4 for $n^\mathrm{h}=7.5$ till $5\times10^{12}$\,cm$^{-2}$, respectively (see black dots in Fig.~\ref{Fig.5}a). 
To establish this result, we particularly focused on the specific shape of the high-field oscillations, as shown in Fig.~\ref{Fig.4}. 
Although a shift of $-2, +2, +4, ...$ on the ratio $E_z/E_c$ gives the same apparent splitting between the $\ket{\mathrm{K} \uparrow}$ and $\ket{\mathrm{K'} \downarrow}$ LLs (see Fig.~\ref{Fig.1}b), the simulated SdH oscillation pattern turns out different at high field. Indeed, the amplitude of the conductivity depends on the LL index \cite{Ando1974} in Eq.~\eqref{eq.4}. Therefore, the relative amplitude of the resistance peaks corresponding to the crossings of $\ket{\mathrm{K} \uparrow}$ and $\ket{\mathrm{K'} \downarrow}$ LLs with the chemical potential varies according to the exact value of $E_z/E_c$. 
Moreover, the system is expected to undergo a transition from a mixed to a polarized LL regime at a critical field depending on $E_z/E_c$.
In this study, no hint of such a transition could be clearly identified. Nevertheless, it is possible to set a higher limit on $E_z/E_c$. For instance, for $V_g=-97.5\,$V (see Supplemental Material II), we observe a low-amplitude resistance peak at 49\,T which we attribute to the crossing of the chemical potential with a $\ket{\mathrm{K'} \downarrow}$ LL.
For this gate voltage, the best fit gave $E_z/E_c = 3.58$. If this ratio was offset by $+2$, an additional large amplitude peak corresponding to a $\ket{\mathrm{K} \uparrow}$ LL should be observed instead, indicating that the transition would occur within our experimental magnetic field range.
A similar conclusion can be drawn for all gate voltages $\ge -100$\,V, ruling out values of $E_z/E_c$ above $\sim 4$ in the explored range of hole densities.

Figure~\ref{Fig.5}a compares our results to those of Movva \textit{et al.} \cite{Movva2017} and Gustafsson \textit{et al.} \cite{Gustafsson2018} marked by gray and blue dots, respectively. In the first study, the $E_z/E_c$ ratio takes integer values only, since the authors detected the magneto-resistance minima at even(odd)-integer values of the filling factor corresponding to odd(even)-integer $E_z/E_c$ ratio \cite{dataMovva}. In the second study, however, the $E_z/E_c$ ratio is determined from the chemical potential jumps between successive LLs, detected using a single-electron transistor coupled to the WSe$_2$ monolayer. Interestingly, the authors assumed a LL structure based on a Schr\"odinger Fermion (SCHF) model. Compared to the massive Dirac Fermion (MDF) model and regarding transport properties only, the main difference lies in the double spin/valley degeneracy of the 0th Landau level for $E_z=0$. In systems where the Zeeman energy is very weak compared to the LL gap, this difference translates into a shift of the SdH oscillations by half a period. This effect has been particularly put forward in graphene to distinguish its peculiar electronic properties (originating from massless Dirac fermions) from those of standard semiconducting 2DEG \cite{Novoselov2005, Zhang2005}. However, for systems such as TMDCs where $E_z$ can be several times larger than $E_c$, the distinction between the two models is uncertain. Indeed, our experimental data could be fitted in a similar way considering the SCHF model with the ratio $E_z/E_c$ up-shifted by 1 (see Figs.~\ref{Fig.5}b and \ref{Fig.5}c). Although our study provides the value of $E_z/E_c$ plus or minus 1 depending on the fermion model, it is noteworthy that its variation versus $n^\mathrm{h}$ shows the same slope compared to the few data available in the literature.

\section{Conclusion}

To conclude, we studied the Zeeman-effect-dependent magneto-resistance oscillation pattern of a WSe$_2$ monolayer sample under strong magnetic field at low temperature. The chemical potential lies in the valence band and the explored range of hole density ranges from $7.5$ till $5\times10^{12}$\,cm$^{-2}$. The high carrier density prevents the observation of the transition from the mixed to the polarized LL regime. 
However, the special shape of the quantum oscillations, which deviates from perfect $1/B$-periodicity, is a clear signature of partially overlapping LLs from the $\ket{\mathrm{K}\uparrow}$ and $\ket{\mathrm{K'}\downarrow}$ states. 
These features are well reproduced using the Gaussian model for conductivity, with no interaction between charge carriers of opposite spins. We extract the ratio $E_z/E_c$, which increases as the hole density decreases, 
thereby enlarging the carrier density range where the enhancement of $g^*$ has been reported.
For systems with negligible Zeeman energy, the magneto-transport technique is sensitive to the degeneracy of the 0th LL and therefore allows a distinction between the LL structure originating from Dirac or Schr\"odinger fermions. 
Here, a large Zeeman energy prevents an unambiguous determination of $E_z/E_c$, as the fermion type in TMDC monolayers is still controversial. In this case, this ratio is inferred  $\pm$1 depending on the fermion model, but cannot be larger than $\sim 5$ in the present study and within the experimentally accessible range of hole density.
The ratio $E_z/E_c$ is directly linked to the effective $g$-factor $g^*=\frac{2m_e}{m^*}\times\frac{E_z}{E_c}$, which ranges between 14.7 and 17.3 for $3.3<E_z/E_c<3.9$ and from 19.1 to 21.8 for $4.3<E_z/E_c<4.9$, considering a fixed effective mass ($m^*=0.45\times m_e$) of the holes.

\begin{acknowledgments}

We acknowledge funding from ANR under project MoS$_2$ValleyControl No.~ANR-14-CE26-0017 and financial support through the EUR grant NanoX No.~ANR-17-EURE-0009 in the framework of the ``Programme des Investissements d'Avenir''. The high magnetic field measurements were performed at LNCMI-Toulouse under EMFL proposals TSC01-118 and TSC07-218.  
We thank Emmanuel Courtade, C\'{e}dric Robert, and their colleagues at LPCNO and Pascal Puech (CEMES) for performing room-temperature photoluminescence characterization of our samples; Thomas Blon (LPCNO) for providing access to a metallic thin-film deposition machine; and our technical staff colleagues at LNCMI-Toulouse.

\end{acknowledgments}


\newpage

\begin{thebibliography}{99}


\bibitem{Splendiani2010}
A.~Splendiani, L.~Sun, Y.~Zhang, T.~Li, J.~Kim, C.-Y. Chim, G.~Galli, and   F.~Wang,
\newblock \href{https://doi.org/10.1021/nl903868w}{Nano Letters \textbf{10}, 1271 (2010)}.

\bibitem{Mak2010}
K.~F. Mak, C.~Lee, J.~Hone, J.~Shan, and T.~F. Heinz,
\newblock \href{https://doi.org/10.1103/PhysRevLett.105.136805}{Physical Review Letters \textbf{105}, 136805 (2010)}.

\bibitem{Yin2012}
Z.~Yin, H.~Li, H.~Li, L.~Jiang, Y.~Shi, Y.~Sun, G.~Lu, Q.~Zhang, X.~Chen, and  H.~Zhang,
\newblock \href{https://doi.org/10.1021/nn2024557}{ACS Nano \textbf{6}, 74  (2012)}.

\bibitem{Lopez-Sanchez2013}
O.~Lopez-Sanchez, D.~Lembke, M.~Kayci, A.~Radenovic, and A.~Kis,
\newblock \href{https://doi.org/10.1038/nnano.2013.100}{Nature  Nanotechnology \textbf{8}, 497 (2013)}.

\bibitem{Xiao2012}
D.~Xiao, G.~B. Liu, W.~X. Feng, X.~D. Xu, and W.~Yao,
\newblock \href{https://doi.org/10.1103/Physrevlett.108.196802}{Physical Review Letters \textbf{108}, 196802 (2012)}.

\bibitem{Zhao2013}
W.~Zhao, Z.~Ghorannevis, L.~Chu, M.~Toh, C.~Kloc, P.-H. Tan, and G.~Eda,
\newblock \href{https://doi.org/10.1021/nn305275h}{ACS Nano \textbf{7}, 791  (2013)}.

\bibitem{Pisoni2018b}
R.~Pisoni, A.~Kormanyos, M.~Brooks, Z.~J. Lei, P.~Back, M.~Eich, H.~Overweg,  Y.~Lee, P.~Rickhaus, K.~Watanabe, T.~Taniguchi, A.~Imamoglu, G.~Burkard, T.~Ihn, and K.~Ensslin,
\newblock \href{https://doi.org/10.1103/Physrevlett.121.247701}{Physical Review Letters \textbf{121}, 247701 (2018)}.

\bibitem{Gustafsson2018}
M.~V. Gustafsson, M.~Yankowitz, C.~Forsythe, D.~Rhodes, K.~Watanabe,  T.~Taniguchi, J.~Hone, X.~Y. Zhu, and C.~R. Dean,
\newblock \href{https://doi.org/10.1038/s41563-018-0036-2}{Nature  Materials \textbf{17}, 411 (2018)}.

\bibitem{Movva2017}
H.~C.~P. Movva, B.~Fallahazad, K.~Kim, S.~Larentis, T.~Taniguchi, K.~Watanabe, S.~K. Banerjee, and E.~Tutuc,
\newblock \href{https://doi.org/10.1103/Physrevlett.118.247701}{Physical Review Letters \textbf{118}, 247701 (2017)}.

\bibitem{Aivazian2015}
G.~Aivazian, Z.~Gong, A.~M. Jones, R.-L. Chu, J.~Yan, D.~G. Mandrus, C.~Zhang,  D.~Cobden, W.~Yao, and X.~Xu,
\newblock \href{https://doi.org/10.1038/nphys3201}{Nature Physics \textbf{11}, 148 (2015)}.

\bibitem{Mitioglu2015}
A.~A. Mitioglu, P.~Plochocka, \'{A}.~Granados~del Aguila, P.~C.~M. Christianen,  G.~Deligeorgis, S.~Anghel, L.~Kulyuk, and D.~K. Maude,
\newblock \href{https://doi.org/10.1021/acs.nanolett.5b00626}{Nano  Letters \textbf{15}, 4387 (2015)}.

\bibitem{Stier2016}
A.~V. Stier, K.~M. McCreary, B.~T. Jonker, J.~Kono, and S.~A. Crooker,
\newblock \href{https://doi.org/10.1038/ncomms10643}{Nature Communications \textbf{7}, 10643 (2016)}.

\bibitem{Koperski2018}
M.~Koperski, M.~R. Molas, A.~Arora, K.~Nogajewski, M.~Bartos, J.~Wyzula,  D.~Vaclavkova, P.~Kossacki, and M.~Potemski,
\newblock \href{https://doi.org/10.1088/2053-1583/aae14b}{2D Materials \textbf{6}, 015001 (2018)}.

\bibitem{Goryca2019}
M.~Goryca, J.~Li, A.~V. Stier, T.~Taniguchi, K.~Watanabe, E.~Courtade, S.~Shree, C.~Robert, B.~Urbaszek, X.~Marie, and S.~A. Crooker,
\newblock \href{https://doi.org/10.1038/s41467-019-12180-y}{Nature Communications \textbf{10}, 4172 (2019)}.

\bibitem{Zhu2011}
Z.~Y. Zhu, Y.~C. Cheng, and U.~Schwingenschlogl,
\newblock \href{https://doi.org/10.1103/Physrevb.84.153402}{Physical Review B \textbf{84}, 153402 (2011)}.

\bibitem{Lin2019}
J.~Lin, T.~Han, B.~A. Piot, Z.~Wu, S.~Xu, G.~Long, L.~An, P.~Cheung, P.~P. Zheng, P.~Plochocka, X.~Dai, D.~K. Maude, F.~Zhang, and N.~Wang,
\newblock \href{https://doi.org/10.1021/acs.nanolett.8b04731}{Nano Letters \textbf{19}, 1736 (2019)}.

\bibitem{Liu2013}
G.-B. Liu, W.-Y. Shan, Y.~Yao, W.~Yao, and D.~Xiao,
\newblock \href{https://doi.org/10.1103/PhysRevB.88.085433}{Physical  Review B \textbf{88}, 085433 (2013)}.

\bibitem{Le2015}
D.~Le, A.~Barinov, E.~Preciado, M.~Isarraraz, I.~Tanabe, T.~Komesu, C.~Troha,  L.~Bartels, T.~S. Rahman, and P.~A. Dowben,
\newblock \href{https://doi.org/10.1088/0953-8984/27/18/182201}{Journal of Physics: Condensed Matter \textbf{27}, 182201 (2015)}.

\bibitem{ref_fabricant}
The WSe$_2$ bulk crystal has been bought from 2Dsemiconductors. (\href{www.2dsemiconductors.com}{www.2dsemiconductors.com})

\bibitem{Castellanos_Gomez_2014}
A.~Castellanos-Gomez, M.~Buscema, R.~Molenaar, V.~Singh, L.~Janssen, H.~S.~J. van~der Zant, and G.~A. Steele,
\newblock \href{https://doi.org/10.1088/2053-1583/1/1/011002}{2D Materials \textbf{1}, 011002 (2014)}.

\bibitem{Rose2013}
F.~Rose, M.~O. Goerbig, and F.~Piechon,
\newblock \href{https://doi.org/10.1103/Physrevb.88.125438}{Physical  Review B \textbf{88}, 125438 (2013)}.

\bibitem{Li2013}
X.~Li, F.~Zhang, and Q.~Niu,
\newblock \href{https://doi.org/10.1103/Physrevlett.110.066803}{ Physical Review Letters \textbf{110}, 066803 (2013)}.

\bibitem{Kormanyos2015}
A.~Kormanyos, P.~Rakyta, and G.~Burkard,
\newblock \href{https://doi.org/10.1088/1367-2630/17/10/103006}{New Journal of Physics \textbf{17}, 103006 (2015)}.

\bibitem{Wang2017}
Z.~F. Wang, J.~Shan, and K.~F. Mak,
\newblock \href{https://doi.org/10.1038/nnano.2016.213}{Nature  Nanotechnology \textbf{12}, 144 (2017)}.

\bibitem{Ando1974}
T.~Ando and Y.~Uemura,
\newblock \href{https://doi.org/10.1143/JPSJ.36.959}{Journal of the  Physical Society of Japan \textbf{36}, 959 (1974)}.

\bibitem{Gerhardts2008}
R.~R. Gerhardts,
\newblock \href{https://doi.org/10.1002/pssb.200743344}{Physica Status Solidi B  \textbf{245}, 378 (2008)}.

\bibitem{Pisoni2018a}
R.~Pisoni, Z.~J. Lei, P.~Back, M.~Eich, H.~Overweg, Y.~Lee, K.~Watanabe,  T.~Taniguchi, T.~Ihn, and K.~Ensslin,
\newblock \href{https://doi.org/10.1063/1.5021113}{Applied Physics  Letters \textbf{112}, 123101 (2018)}.

\bibitem{dataMovva}
Only part of the results from the original publication has been reproduced. An  alternative set of data would be the same plot with the $E_z/E_c$ values  downshifted by 2.

\bibitem{Novoselov2005}
K.~S. Novoselov, A.~K. Geim, S.~V. Morozov, D.~Jiang, M.~I. Katsnelson, I.~V. Grigorieva, S.~V. Dubonos, and A.~A. Firsov,
\newblock \href{https://doi.org/10.1038/nature04233}{Nature \textbf{438}, 197  (2005)}.

\bibitem{Zhang2005}
Y.~Zhang, Y.-W. Tan, H.~L. Stormer, and P.~Kim,
\newblock \href{https://doi.org/10.1038/nature04235}{Nature \textbf{438}, 201  (2005)}.

\end{thebibliography}
\end{document}


\title{High magnetic field spin-valley-split Shubnikov--de Haas oscillations in a WSe$_2$ monolayer\\ \bigskip Supplemental Material \bigskip}

\author{Banan Kerdi\,\orcidicon{0000-0002-6015-1850}}
\email{banan.kerdi@lncmi.cnrs.fr}
\affiliation{LNCMI, Universit\'e de Toulouse, CNRS, INSA, UPS, EMFL, 31400 Toulouse, France}

\author{Mathieu Pierre\,\orcidicon{0000-0001-5274-4309}}
\affiliation{LNCMI, Universit\'e de Toulouse, CNRS, INSA, UPS, EMFL, 31400 Toulouse, France}

\author{Robin Cours}
\affiliation{CEMES, Universit\'e de Toulouse, CNRS, 31055 Toulouse, France}

\author{B\'{e}n\'{e}dicte Warot-Fonrose\,\orcidicon{0000-0001-9829-9431}}
\affiliation{CEMES, Universit\'e de Toulouse, CNRS, 31055 Toulouse, France}

\author{Michel Goiran}
\affiliation{LNCMI, Universit\'e de Toulouse, CNRS, INSA, UPS, EMFL, 31400 Toulouse, France}

\author{Walter Escoffier}
\email{walter.escoffier@lncmi.cnrs.fr}
\affiliation{LNCMI, Universit\'e de Toulouse, CNRS, INSA, UPS, EMFL, 31400 Toulouse, France}

\maketitle

\vspace{3cm}

\section{Magneto-resistance oscillations versus filling factor}

\begin{figure}[!htp]
\begin{center}
\includegraphics[width=0.8\linewidth]{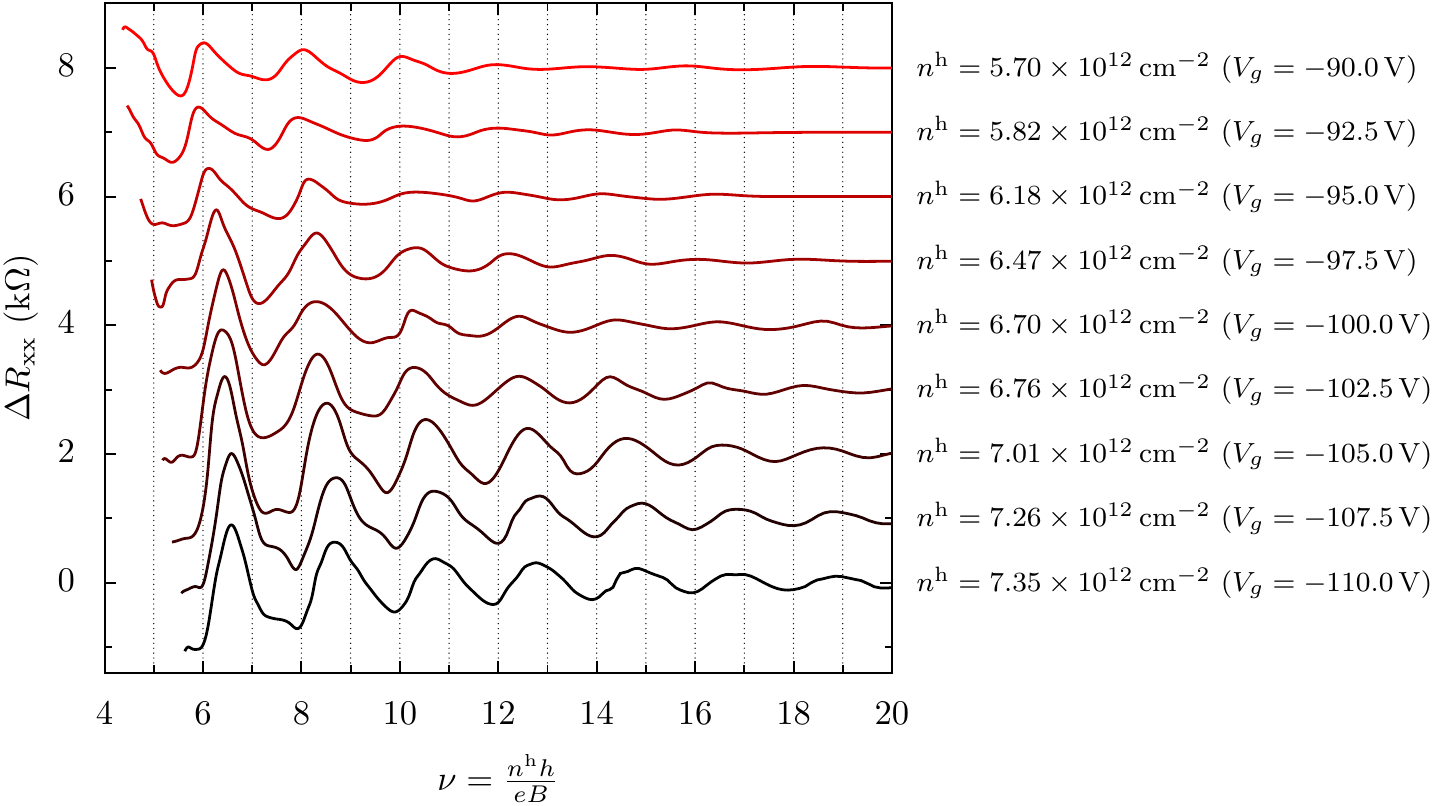}
\caption{Experimental magneto-resistance oscillations (background subtracted) at $T=4.2$\,K plotted versus filling factor $\nu=\frac{n^\mathrm{h}h}{eB}$. The curves are successively offset by 1\,k$\Omega$ for clarity. The carrier density was extracted from the fitting procedure described in the main text.}
\end{center}
\end{figure}

\newpage

\section{Additional fitting of the experimental magneto-resistance oscillations}
\vspace{-0.5cm}

\begin{figure}[!h]
\begin{center}
\includegraphics[width=1.0\linewidth]{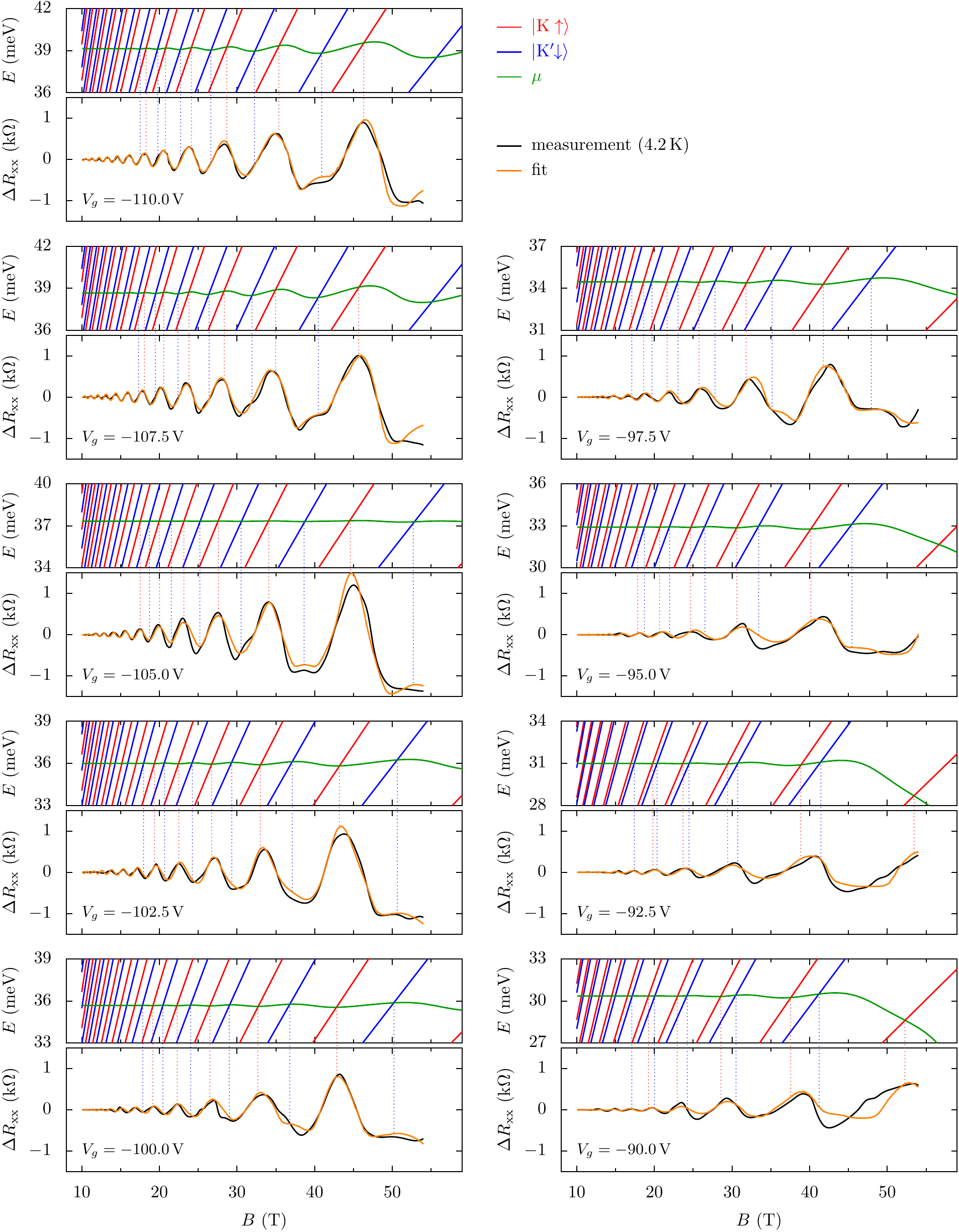}
\caption{Fit of the magneto-resistance oscillations and corresponding LL structure and chemical potential for all measurements at gate voltages between $-110$ and $-90$\,V. The fitting parameters are given in Table~\ref{table:allfits}.\label{fig:allfits}}
\end{center}
\end{figure}

\newpage

~\vspace{2cm}

\begin{table}[!htp]
\renewcommand{\arraystretch}{1.5}
\begin{tabular}{|  m{3cm}   || *{9}{>{\centering\arraybackslash}m{1.2cm}|}} 
	\hline
   $V_g$ (V) & $-110$ & $-107.5$ & $-105$ & $-102.5$ & $-100$ & $-97.5$ &  $-95$ & $-92.5$ & $-90$ \\ \hline\hline
   $n^\text{h}$ $\left(\times 10^{12}\,\text{cm}^{-2}\right)$  & 7.35 &   7.26 & 7.01 &   6.76 & 6.70 &  6.47 & 6.18  & 5.82  & 5.70  \\ \hline
   $E_z/E_c$  &3.37 &   3.36 & 3.49 &   3.56 & 3.55 &  3.62 & 3.66 &  3.83 & 3.75  \\ \hline
   $\mu^\text{h}$ $\left(\text{cm}^{2}\text{V}^{-1}\text{s}^{-1}\right)$  & 2400 & 2400 & 2400 & 2400 & 2200 & 1600 & 1400 & 1200 & 1300  \\ \hline
\end{tabular}
\caption{Hole density, $E_z/E_c$ ratio, and zero-field mobility obtained from the fit of the experimental quantum oscillations shown in Fig.~\ref{fig:allfits}.\label{table:allfits}}
\end{table}

\vspace{3cm}

\section{Comparison of the hole carrier density extracted from the fit and the plane capacitor model}

\begin{figure}[!htp]
\begin{center}
\includegraphics[width=.5\linewidth]{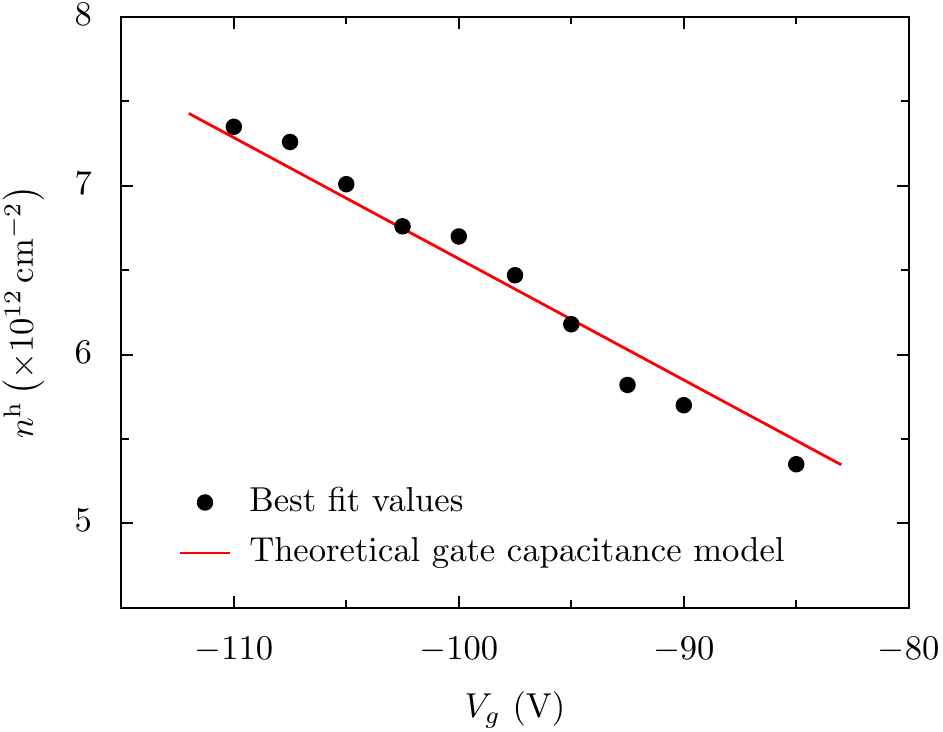}
\caption{The hole carrier density extracted from the fitting of the magneto-resistance oscillations is compared to its theoretical evolution with back-gate voltage, based on the calculation of the back-gate capacitance with the plane capacitor model, which gives $C_g = 11.5\,\mathrm{nF}/\mathrm{cm}^2$.}
\end{center}
\end{figure}

\newpage

\section{Determination of the cyclotron mass}

\begin{figure}[!h]
\begin{center}
\includegraphics[width=.8\linewidth]{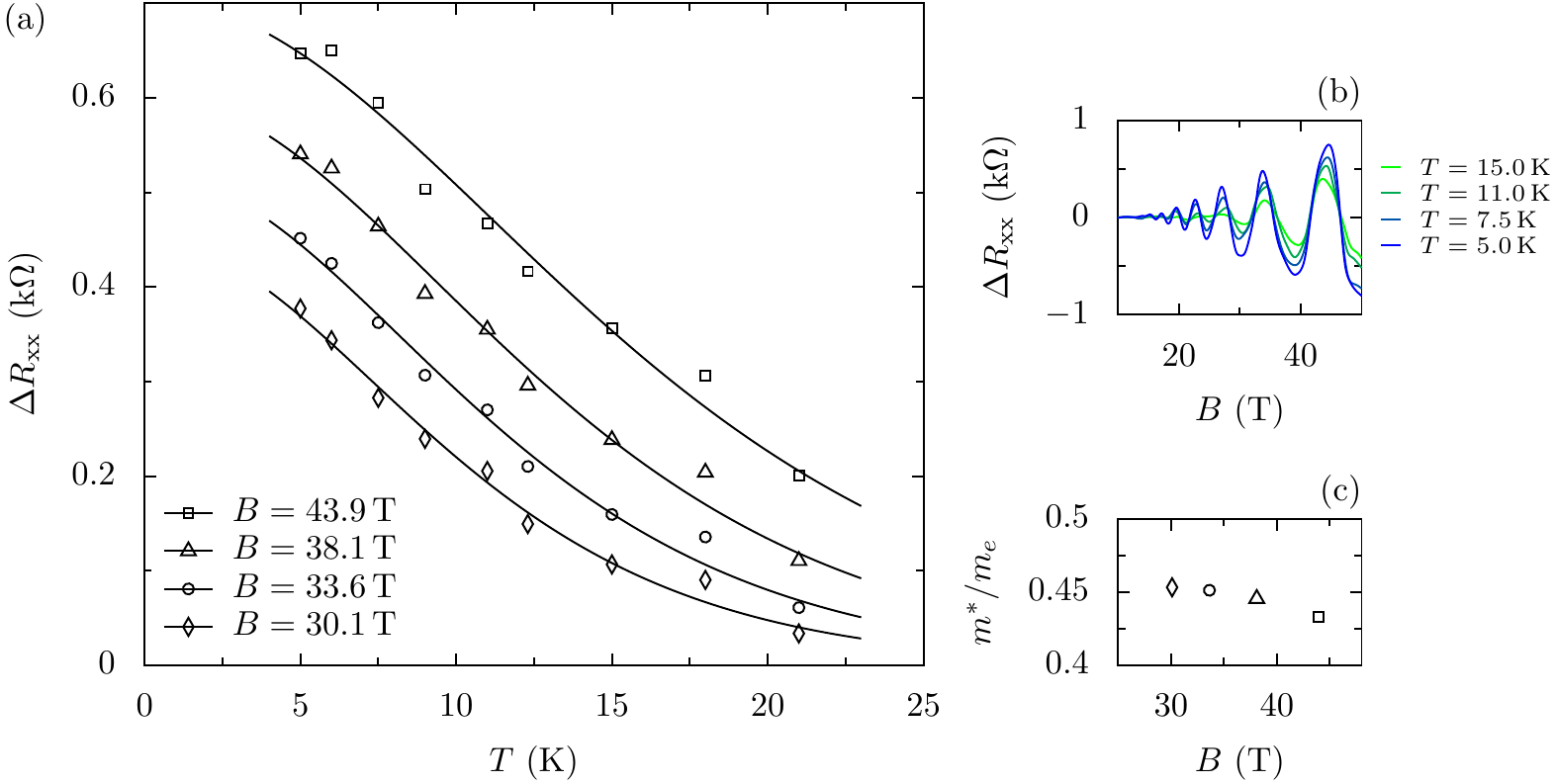}
\caption{(a) Amplitudes of the four highest magnetic field oscillations measured at for $V_g=-105$\,V versus temperature. Their decay with temperature is fitted with the Lifshitz-Kosevich model: $\Delta R_\mathrm{xx}(T) \propto x/\sinh(x)$ with $x = (2 \pi^2m^*k_\text{B}T)/(\hbar e B)$. (b)~Corresponding magneto-resistance oscillation measurements (background subtracted) for selected temperatures. (c) Effective mass resulting from the fits shown in panel (a). We extract an effective mass $m^* = (0.45 \pm 0.04) \times m_e$, which confirms previous reports in the literature for WSe$_2$ monolayers.}
\end{center}
\end{figure}